\pdfoutput=1

\documentclass[twocolumn,superscriptaddress,aps,preprintnumbers,amsmath,amssymb,prl,nofootinbib]{revtex4}

\usepackage{epsfig}
\usepackage{amsmath}
\usepackage{amssymb}
\usepackage{subfigure}
\usepackage{cancel}
\usepackage{textcomp}
\usepackage{calc}
\usepackage{graphicx}
\usepackage{dcolumn}
\usepackage{color}
\usepackage{xcolor}
\usepackage{pifont}
\usepackage{slashed}

\usepackage{hyperref}
\hypersetup{colorlinks=true,urlcolor=blue,linkcolor=red,citecolor=green!60!black}

\newcommand\ii{\mathrm{i}}

\begin{document}


\preprint{CERN-TH-2020-055}

\title{Fingerprint of Low-Scale Leptogenesis in the Primordial Gravitational-Wave Spectrum}

\author{Simone Blasi}
\email{blasi@mpi-hd.mpg.de}
\affiliation{Max-Planck-Institut  f{\"u}r  Kernphysik,  69117  Heidelberg,  Germany}

\author{Vedran Brdar}
\email{vbrdar@mpi-hd.mpg.de}
\affiliation{Max-Planck-Institut  f{\"u}r  Kernphysik,  69117  Heidelberg,  Germany}

\author{Kai Schmitz}
\email{kai.schmitz@cern.ch}
\affiliation{Theoretical Physics Department, CERN, 1211 Geneva 23, Switzerland}


\begin{abstract}
The dynamical generation of right-handed-neutrino (RHN) masses in the early Universe naturally entails the formation of cosmic strings that give rise to an observable signal in gravitational waves (GWs).
Here, we show that a characteristic break in the GW spectrum would provide evidence for a new stage in the cosmological expansion history and a suppression of the RHN mass scale compared to the scale of spontaneous symmetry breaking.
The detection of such a spectral feature would thus represent a novel and unique possibility to probe the physics of RHN mass generation in regions of parameter space that allow for low-scale leptogenesis in accord with electroweak naturalness.
\end{abstract}


\date{\today}
\maketitle


\noindent\textbf{Introduction.}
The seesaw mechanism~\cite{Minkowski:1977sc,Yanagida:1979as,Yanagida:1980xy,GellMann:1980vs,Mohapatra:1979ia} resolves two puzzles of physics beyond the Standard Model (SM) at the same time: neutrino oscillations~\cite{Tanabashi:2018oca} and the baryon asymmetry of the Universe~\cite{Aghanim:2018eyx}.
At its core, it is based on an extension of the SM by a set of right-handed neutrinos (RHNs) $N_i$ that do not carry any SM gauge charge and that possess (potentially very large) lepton-number-violating Majorana masses $M_i$.
These sterile neutrinos induce small Majorana masses for the active SM neutrinos, which explains their flavor oscillations, while RHN decays or oscillations in the early Universe can create a primordial baryon asymmetry via leptogenesis~\cite{Fukugita:1986hr,Akhmedov:1998qx}.


The Majorana masses $M_i$ are free input parameters of the seesaw mechanism, which raises the question of their origin at high energies.
An attractive ultraviolet completion of the seesaw model consists in promoting the difference of baryon number $B$ and lepton number $L$, which is an accidental global symmetry of the SM, to a new gauge symmetry, $U(1)_{B-L}$~\cite{Davidson:1978pm,Marshak:1979fm,Mohapatra:1980qe}, such that the RHNs acquire their mass in consequence of the spontaneous breaking of this symmetry.
This scenario also sets the stage for embedding the seesaw mechanism in grand unified theories (GUTs) that feature the product of $U(1)_{B-L}$ and the SM gauge group $G_{\rm SM} = SU(3)_C \times SU(2)_L \times U(1)_Y$ as a subgroup of the GUT gauge group, $G_{\rm GUT} \supset G_{\rm SM} \times U(1)_{B-L}$.


Most of the seesaw parameter space is hard to test in terrestrial experiments because the RHNs are either too heavy or too weakly coupled~\cite{Chrzaszcz:2019inj}.
It is therefore remarkable that a high $B\!-\!L$ breaking scale can still be probed by a different observable\,---\,primordial gravitational waves (GWs)~\cite{Caprini:2018mtu,Christensen:2018iqi}.
In recent years, it has been realized that the spontaneous breaking of $U(1)_{B-L}$ during a cosmological phase transition in the early Universe easily results in a strong GW signal.
This was first demonstrated in~\cite{Buchmuller:2013lra} (see \cite{Buchmuller:2019gfy} for an update), which considered a second-order $B\!-\!L$ phase transition after cosmic inflation~\cite{Buchmuller:2010yy,Buchmuller:2011mw,Buchmuller:2012wn,Buchmuller:2012bt,Schmitz:2012kaa,Buchmuller:2013dja,Domcke:2017xvu,Domcke:2017rzu}, and more recently revisited in a more general context in~\cite{Dror:2019syi}.
The scenarios described in~\cite{Buchmuller:2013lra,Buchmuller:2019gfy,Dror:2019syi} share the common property that the spontaneous breaking of $U(1)_{B-L}$ results in a network of local cosmic strings that emit a large stochastic GW background~\cite{Vachaspati:1984gt} (see \cite{Auclair:2019wcv} for a review).
Alternatively, the $B\!-\!L$ phase transition itself can result in an observable GW signal if it is of first order~\cite{Chao:2017ilw,Okada:2018xdh,Hasegawa:2019amx,Haba:2019qol} and especially in the classically conformal limit~\cite{Jinno:2016knw,Iso:2017uuu,Marzo:2018nov,Hashino:2018wee,Bian:2019szo}.
In this case, future GW experiments will be able to probe the $B\!-\!L$ breaking scale up to $v_{B-L} \sim 10^9\,\textrm{GeV}$, assuming a strongly supercooled phase transition~\cite{Jinno:2016knw,Hashino:2018wee}.
Meanwhile, the complementary range all the way up to the unification scale, $10^9\,\textrm{GeV} \lesssim v_{B-L} \lesssim 10^{16}\,\textrm{GeV}$, is expected to yield a strong GW signal from cosmic strings~\cite{Dror:2019syi}.


A potential drawback of high-scale $B\!-\!L$ breaking is that it may aggravate the SM hierarchy problem.
Indeed, RHN threshold corrections to the mass of the SM Higgs boson $h$ spoil electroweak (EW) naturalness for RHN masses larger than $M_i \sim 10^7\,\textrm{GeV}$~\cite{Vissani:1997ys,Clarke:2015gwa} (see also~\cite{Brivio:2017dfq,Brivio:2018rzm,Brdar:2019iem,Brivio:2019hrj}).
In absence of any additional cancellation mechanism, there are two ways out of this problem:
(i) Regarding $v_{B-L}$ as an independent input scale, one may simply assume it to be small enough, so that $M_i \lesssim 10^7\,\textrm{GeV}$ for all RHNs.
(ii) Insisting on a large $v_{B-L}$ value, as motivated by grand unification, one may assume a parametric suppression of the RHN mass scale compared to the scale of $B\!-\!L$ breaking by means of small Yukawa couplings.


In this paper, we will follow the second approach and scrutinize the resulting cosmic-string-induced GW signal.
In particular, we will argue that the detection of a characteristic break in the GW spectrum would point to a new stage in the cosmological expansion history and hence provide evidence for high-scale $B\!-\!L$ breaking, $v_{B-L} \gtrsim 10^9\,\textrm{GeV}$, together with light RHN masses in accord with EW naturalness, $M_i \lesssim 10^7\,\textrm{GeV}$.
An important outcome of our analysis is that the GW spectral index is expected to change from $n_{\rm gw} \simeq 0$ to $n_{\rm gw} \simeq -1/3$, which deviates from earlier results in the literature and which holds for a broad class of modified early-Universe scenarios, including nonstandard matter domination.
In addition, we study a concrete and minimal particle physics model, the minimal gauged $B\!-\!L$ model, which already contains all the necessary ingredients.
For this model, we show that the break in the GW spectrum encodes information\,---\,not only on the $B\!-\!L$ breaking scale\,---\,but also on the RHN mass scale.
Future GW experiments will thus be able to probe RHN masses relevant for leptogenesis at intermediate and low energies (see, e.g.,~\cite{Akhmedov:1998qx,Pilaftsis:1997jf,Pilaftsis:2003gt,Boubekeur:2004ez,Raidal:2004vt,Moffat:2018wke,Hugle:2018qbw,Baumholzer:2018sfb,Borah:2018rca,Abada:2018oly,Alanne:2018brf,Mahanta:2019gfe,Baumholzer:2019twf}).


\noindent\textbf{Break in the gravitational-wave spectrum.}
We begin with a model-independent argument based on two assumptions:
(i) The symmetry-breaking phase transition is of second order and (ii) the symmetry-breaking scalar field $\phi$ has a long lifetime $t_\phi$.
In this case, $\phi$ will coherently oscillate for a long time after the phase transition around the true, symmetry-breaking vacuum, and the energy density stored in these oscillations, $\rho_\phi$, will redshift like pressureless dust.
$\rho_\phi$ will in particular be diluted less fast than the energy density of the radiation background, $\rho_{\rm rad}$, such that it may become the dominant form of energy in the Universe at some time $t < t_\phi$.
As we will show below, such a nonstandard era of matter domination, or \textit{scalar era}, results in a characteristic break in the cosmic-string-induced GW spectrum.
The observation of this break would thus allow one to reconstruct the expansion history of the Universe and infer the presence of a long-lived, i.e., weakly coupled scalar field (see~\cite{Cui:2017ufi,Cui:2018rwi,Caldwell:2018giq,DEramo:2019tit,Bernal:2019lpc,Figueroa:2019paj,Gouttenoire:2019kij,Gouttenoire:2019rtn} for related work on using GWs for \textit{cosmic archaeology}).
This in turn would point toward a parametric separation between the scale of symmetry breaking (which must be large, otherwise we would not observe any signal) and the mass scale of the particles that become massive during the phase transition (which must be small, otherwise $\phi$ would be too short-lived).
In the context of the seesaw model, the break in the GW spectrum would therefore be a \textit{sufficient} indication allowing one to claim a low RHN mass scale in agreement with EW naturalness and low-scale leptogenesis.
Meanwhile, it does not constitute a \textit{necessary} requirement.
If no break should be detected, the RHN mass scale may or may not be suppressed.
In this case, one would simply have to assume that the field $\phi$ possesses at least one fast decay channel.


Let us now study the GW signal in more detail.
We will follow~\cite{Auclair:2019wcv,Cui:2017ufi,Cui:2018rwi} and compute the GW energy density spectrum $\Omega_{\rm gw}$ based on the analytic velocity-dependent one-scale model for cosmic strings~\cite{Martins:1995tg,Martins:1996jp,Martins:2000cs,Sousa:2013aaa} (see also~\cite{Sousa:2020sxs}),
\begin{equation}
\label{eq:OGW}
\Omega_{\rm gw}\left(f\right) = \sum_{k=1}^{\infty} \Omega_{\rm gw}^{(k)}\left(f\right) = \frac{8\pi}{3H_0^2} \left(G\mu\right)^2 f \sum_{k=1}^{\infty} C_k P_k \,.
\end{equation}
Here, $H_0 \simeq 67\,\textrm{km}/\textrm{s}/\textrm{Mpc}$~\cite{Aghanim:2018eyx} is the current Hubble rate; $G$ is Newton's constant; $\mu$ is the cosmic-string tension; $k$ counts the harmonic excitations of cosmic-string loops; and $P_k = \Gamma/k^{q}/\zeta\left(q\right)$ is the corresponding averaged power spectrum for GWs emitted by cusps propagating along cosmic-string loops ($q = 4/3$).
The normalization of $P_k$ is fixed by the total emitted power $\Gamma = \sum_k P_k$, which follows from numerical simulations, $\Gamma \simeq 50$~\cite{Blanco-Pillado:2013qja,Blanco-Pillado:2017oxo}.
The function $C_k$ represents an integral from the onset of the cosmic-string scaling regime, $t_{\rm scl}$, to the present time $t_0$,
\begin{equation}
C_k\left(f\right) = \frac{2k}{f^2}\int_{t_{\rm scl}}^{t_0}dt\:\theta\left(t_k-t_{\rm scl}\right)\left(\frac{a\left(t_{\vphantom{0}}\right)}{a\left(t_0\right)}\right)^5 n\left(\ell_k,t\right) \,,
\end{equation}
where $a$ denotes the cosmic scale factor; $t_k$ is the time when the loops that contribute to the present-day GW frequency $f$ via their $k^{\rm th}$ harmonic mode were formed,
\begin{equation}
t_k = \frac{\ell_k/t + \Gamma\,G\mu}{\alpha + \Gamma\,G\mu}\,t \,,\quad \ell_k = \frac{2k}{f} \frac{a\left(t_{\vphantom{0}}\right)}{a\left(t_0\right)} \,;
\end{equation}
and $n$ is the number of loops per volume and unit length,
\begin{equation}
n\left(\ell_k,t\right) = \frac{F}{t_k^4}\left(\frac{a\left(t_k\right)}{a\left(t_{\vphantom{0}}\right)}\right)^3 \frac{C_{\rm eff}\left(t_k\right)}{\alpha\left(\alpha + \Gamma\,G\mu\right)} \,.
\end{equation}
$F = 0.1$ is an efficiency factor~\cite{Blanco-Pillado:2013qja,Sanidas:2012ee}; $\alpha = \ell_k/t_k$ characterizes the loop size at the time of formation (below, we will use the characteristic value $\alpha = 0.1$); and $C_{\rm eff}$ distinguishes between loops formed during radiation ($C_{\rm eff} \simeq 5.4$) and matter ($C_{\rm eff} \simeq 0.39$) domination.
Below, we will simply switch between these discrete values for $C_{\rm eff}$ whenever the dominant form of energy changes.


\begin{table}
\renewcommand\arraystretch{1.25}
\begin{center}
\caption{Benchmark scenarios highlighted in Figs.~\ref{fig:BPs} and \ref{fig:scan}.
$v_{B-L}$, $m_N$, $m_\phi$, and $\Gamma_\phi$ are stated in units of GeV.}
\label{tab:BPs}
\smallskip
\begin{tabular}{|c||cccccc|}
\hline
    & $g_{B-L}$ & $v_{B-L}$            & $m_N$ & $m_\phi$           & $\Gamma_\phi$       & $G \mu$             \\
\hline\hline
BP1 & $10^{-4}$ & $ 4 \times 10^{13}$  & 2     & $2 \times 10^6$    & $3 \times 10^{-22}$ & $3 \times 10^{-12}$ \\
BP2 & $10^{-3}$ & $ 5 \times 10^{12}$  & 5     & $2 \times 10^7$    & $1 \times 10^{-18}$ & $6 \times 10^{-14}$ \\
BP3 & $10^{-2}$ & $ 3 \times 10^{13}$  & 200   & $2 \times 10^{10}$ & $2 \times 10^{-13}$ & $6 \times 10^{-12}$ \\
\hline
\end{tabular}
\end{center}
\end{table}


Next, let us analyze the shape of the GW spectrum.
The spectrum emitted by the fundamental mode of each cosmic-string loop, $k=1$, in the presence of a nonstandard scalar era was investigated in \cite{Cui:2017ufi,Cui:2018rwi,Auclair:2019wcv,Gouttenoire:2019kij,Gouttenoire:2019rtn},
\begin{equation}
\label{eq:OGW1}
\Omega_{\rm gw}^{(1)}\left(f\right) \sim \Omega_{\rm gw}^{\rm plt} \left(\frac{f}{f_{\rm brk}}\right)^{n_{\rm gw}} \,,
\end{equation}
with $n_{\rm gw} \simeq 0$ for $f \lesssim f_{\rm brk}$ and $n_{\rm gw} \simeq -1$ for $f \gtrsim f_{\rm brk}$.
The spectrum hence features a more or less flat plateau,
\begin{equation}
\Omega_{\rm gw}^{\rm plt} \sim \frac{10^{-3}}{\zeta\left(q\right)} \bigg(\frac{\alpha}{0.1}\bigg)^{1/2}\left(\frac{G\mu}{\Gamma}\right)^{1/2} \,,
\end{equation}
up to a characteristic break in the spectrum located at
\begin{equation}
\label{eq:fbrk}
f_{\rm brk} \simeq \left(\frac{8\,z_{\rm eq}\,t_{\rm eq}}{\alpha\,\Gamma\,G \mu\,t_{\rm end}}\right)^{1/2} \frac{1}{t_0} \,,
\end{equation}
where $t_{\rm eq}$ and $z_{\rm eq}$ denote the redshift and time at standard matter--radiation equality, and where $t_{\rm end}$ marks the end of the scalar era.
In our case, we have $t_{\rm end} \simeq t_\phi \simeq 1/\Gamma_\phi$, with $\Gamma_\phi$ being the decay rate of the scalar field $\phi$.


\begin{figure*}
\centering
\includegraphics[width=0.317\textwidth]{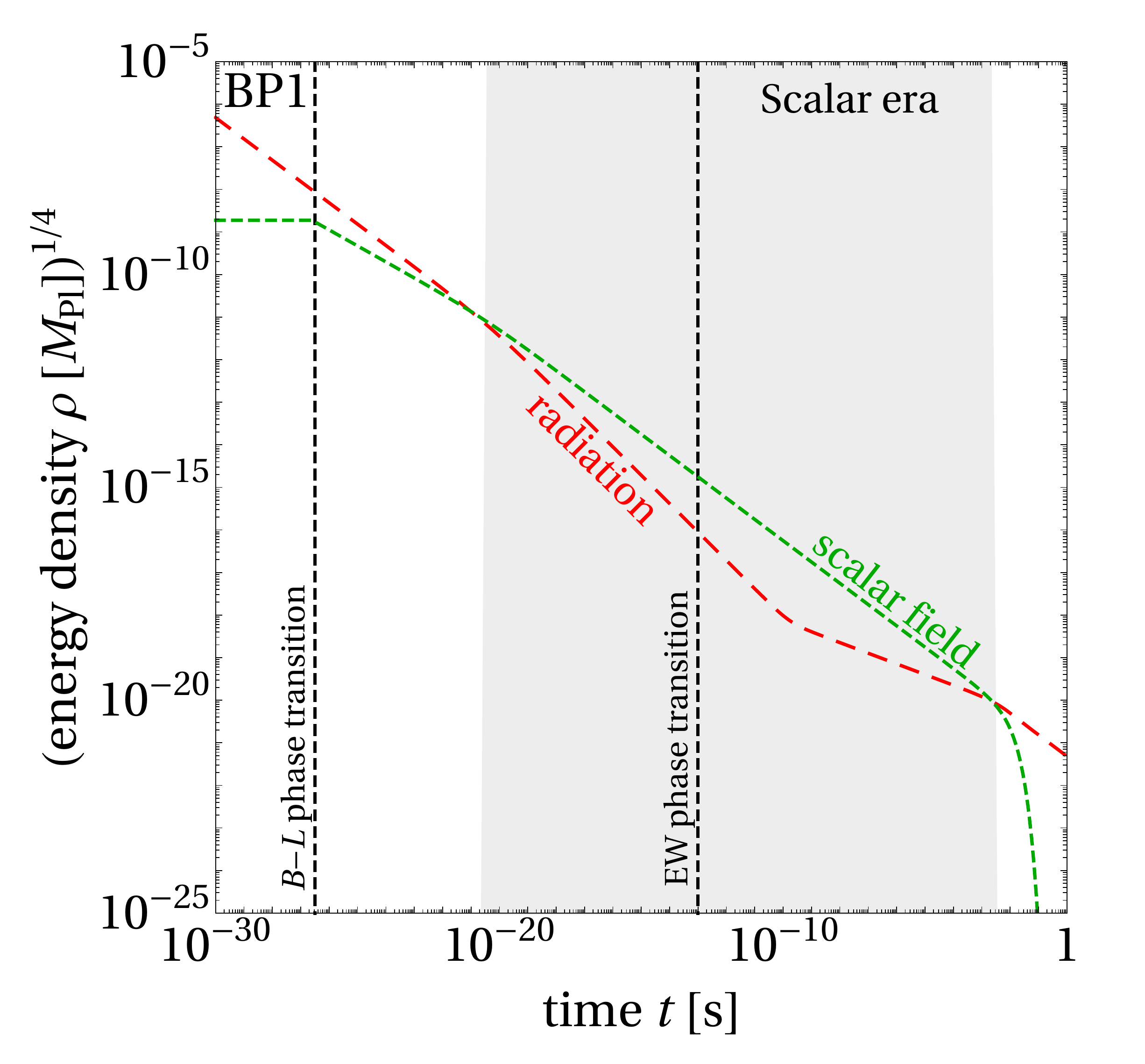}\quad
\includegraphics[width=0.320\textwidth]{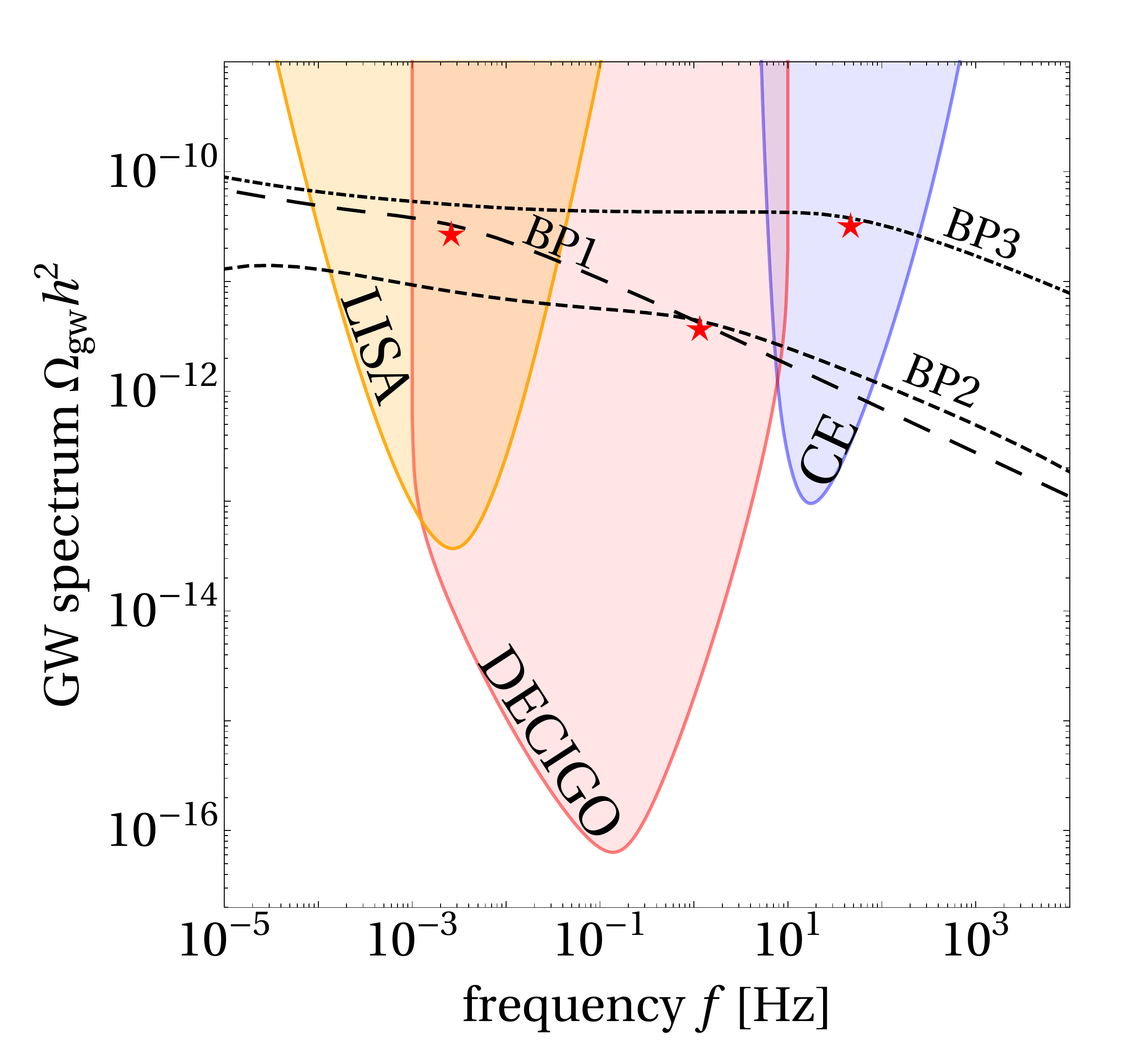}\quad
\includegraphics[width=0.310\textwidth]{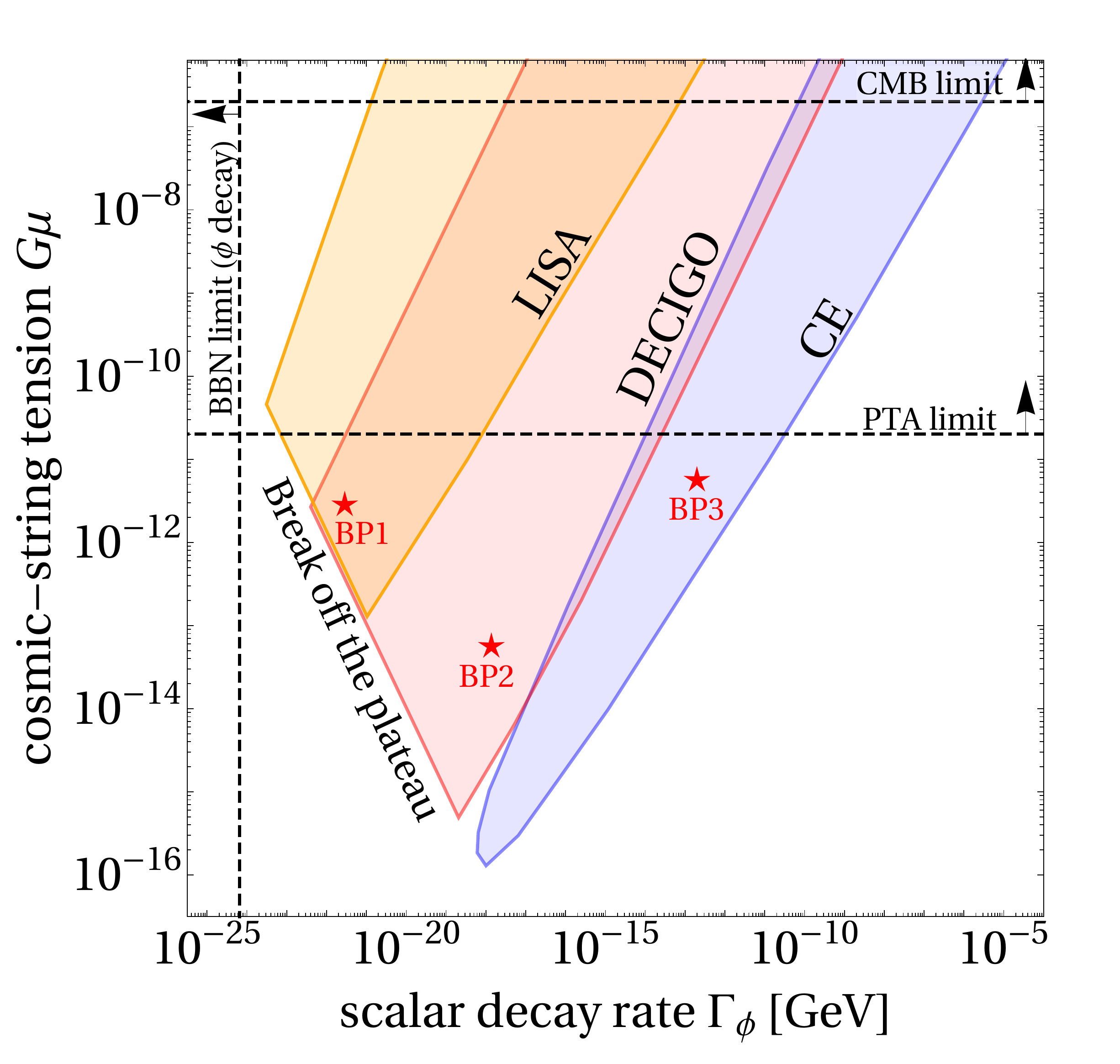}
\caption{Scalar era after the $B\!-\!L$ phase transition.
\textit{Left:} Evolution of the scalar-field and radiation energy densities.
\textit{Middle}: GW spectra and experimental sensitivities.
\textit{Right}: Bounds and projected sensitivities in the $\Gamma_\phi$--\,$G\mu$ plane.
See text and Tab.~\ref{tab:BPs}.}
\label{fig:BPs}
\end{figure*}


Eq.~\eqref{eq:OGW1}, and in particular the GW spectral index $n_{\rm gw}$, are nontrivially modified by the GW emission from the higher cosmic-string modes.
First of all, we note that
\begin{equation}
\Omega_{\rm gw}^{(k)}\left(f\right) = \frac{1}{k^{q}}\:\Omega_{\rm gw}^{(1)}\left(\frac{f}{k}\right) \,.
\end{equation}
Then, if we only sum up the approximately flat parts of the individual spectra, neglecting the $1/f$ tails, we find
\begin{equation}
\Omega_{\rm gw}\left(f\right) \sim \Omega_{\rm gw}^{\rm plt} \sum_{k=m}^\infty \frac{1}{k^{q}} = \Omega_{\rm gw}^{\rm plt} \,\zeta\left(q,m\right) \,, 
\end{equation}
where $m$ is the first integer that is larger than $f/f_{\rm brk}$, and where $\zeta\left(q,m\right)$ denotes the Hurwitz zeta function.
At large frequencies, $f \gg f_{\rm brk}$, we therefore obtain
\begin{equation}
\Omega_{\rm gw}\left(f\right) \sim 3\,\Omega_{\rm gw}^{\rm plt}\left(\frac{f}{f_{\rm brk}}\right)^{-1/3} \left[1 + \mathcal{O}\left(\frac{f_{\rm brk}}{f}\right)\right] \,.
\end{equation}
This is our first main result.
Including the contributions to the GW spectrum from all harmonic modes, one finds that $n_{\rm gw}$ changes in fact from $n_{\rm gw} \simeq 0$ to $n_{\rm gw} \simeq -1/3$ around $f_{\rm brk}$.
%
%
We also argue that our result can be generalized to any $k = 1$ GW spectrum that falls off faster than $f^{-1/3}$ at high frequencies.
In this case, the sum of the individual flat contributions will always represent an irreducible background with a spectral index of $n_{\rm gw} \simeq -1/3$ (see also the discussion on GWs diluted during inflation in~\cite{Guedes:2018afo,Cui:2019kkd}).
This is also illustrated in Fig.~\ref{fig:BPs}, where we compare the GW spectra for three benchmark points (BPs) in the minimal gauged $B\!-\!L$ model (see Tab.~\ref{tab:BPs} and below) based on the full expression in Eq.~\eqref{eq:OGW} with the power-law-integrated sensitivity curves of three future GW experiments~\cite{Thrane:2013oya}: Cosmic Explorer (CE)~\cite{Evans:2016mbw,Reitze:2019iox}, the Deci-Hertz Interferometer Gravitational-Wave Observatory (DECIGO)~\cite{Seto:2001qf,Kawamura:2006up}, and the Laser Interferometer Space Antenna (LISA)~\cite{Audley:2017drz,Baker:2019nia} (see~\cite{Schmitz:2020syl} for details).


In the right panel of Fig.~\ref{fig:BPs}, we indicate where in the $\Gamma_\phi$--\,$G\mu$ plane we expect CE, DECIGO, and LISA to respectively observe a break in the \textit{flat} part of the GW spectrum.
An analysis of nonstandard features in the \textit{nonflat} part, where the GW spectrum is affected by the standard radiation--matter transition in the early Universe, is left for future work.
We also include upper bounds on $G\mu$ from observations of the cosmic microwave background (CMB), $G \mu \lesssim 10^{-7}$~\cite{Ade:2015xua,Charnock:2016nzm,Lizarraga:2016onn}, as well as from pulsar timing array (PTA) data, $G \mu \lesssim 2 \times 10^{-11}$~\cite{Blanco-Pillado:2017rnf,Ringeval:2017eww}.
These bounds only apply when the cosmic strings are topologically stable, which may no longer be the case when $G_{\rm SM}\times U(1)_{B-L}$ is embedded in a semisimple GUT group at high energies~\cite{Vilenkin:1982hm,Monin:2008mp} (see~\cite{Buchmuller:2019gfy,Dror:2019syi} for details).


\noindent\textbf{Minimal gauged \boldmath{$B\!-\!L$} model.}
The scenario described in the previous section is already realized in a minimal SM extension.
To see this, we extend $G_{\rm SM}$ by a $U(1)_{B-L}$ factor and supplement the SM particle content by:
the $B\!-\!L$ vector boson $Z'$ with gauge coupling $g_{B-L}$,
a complex symmetry-breaking scalar field $\Phi = \phi/\sqrt{2}\,e^{\ii\theta}$ with $B\!-\!L$ charge $q_\phi = -2$ and vanishing SM charges,
and three sterile RHNs $N_i$ ($i=1,2,3$) with universal $B\!-\!L$ charge $q_N = -1$.
The new interaction Lagrangian reads
\begin{align}
\label{eq:L}
\Delta\mathcal{L} =
& - \bigg[ y_{i\alpha}^{\rm D}\,\overline{N_i^{\rm R}}\,\tilde{H}^\dagger L_\alpha + \frac{1}{2}\,y_i^{\rm M}\,\Phi\,\overline{N_i^{\rm R}}\left(N_i^{\rm R}\right)^{\rm C} + \textrm{H.c.} \bigg]
\nonumber \\
& - \bigg[ \lambda_\phi\left(\left|\Phi\right|^2 - \frac{1}{2}\,v_{B-L}^2\right)^2 + \lambda_{\phi h}\left|\Phi\right|^2\left|H\right|^2 \bigg] \,.
\end{align}
Below, we will assume $y_i^{\rm M} \simeq y_N$ for all $i$, which translates to $M_i = y_i^{\rm M}/\sqrt{2}\,v_{B-L} \simeq m_N$ after $B\!-\!L$ breaking.
This will simplify our analysis and is consistent with the notion of resonant leptogenesis at low energies~\cite{Pilaftsis:1997jf,Pilaftsis:2003gt}.


\begin{figure*}
\centering
\includegraphics[width=0.3\textwidth]{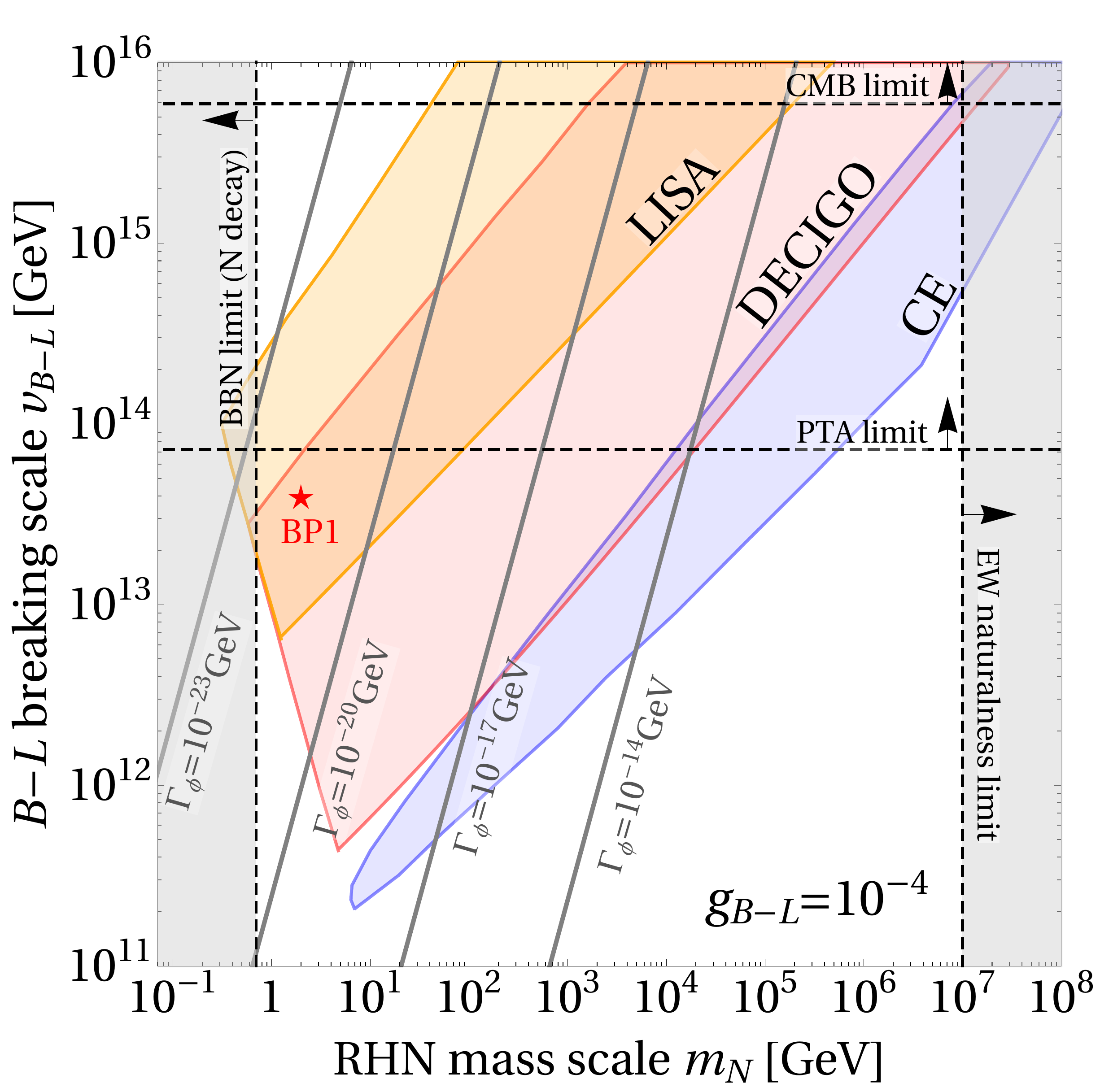}\qquad
\includegraphics[width=0.3\textwidth]{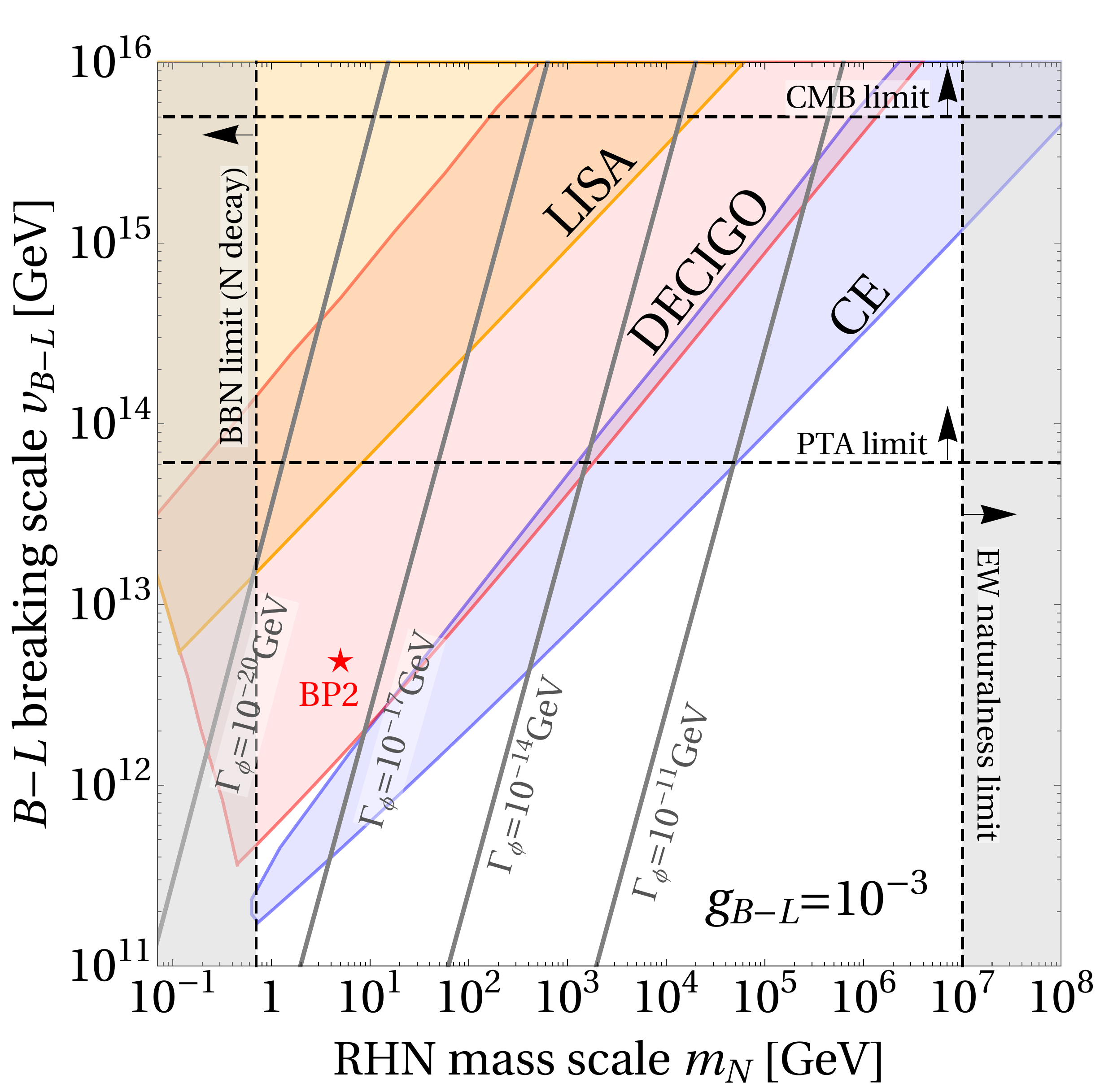}\qquad
\includegraphics[width=0.3\textwidth]{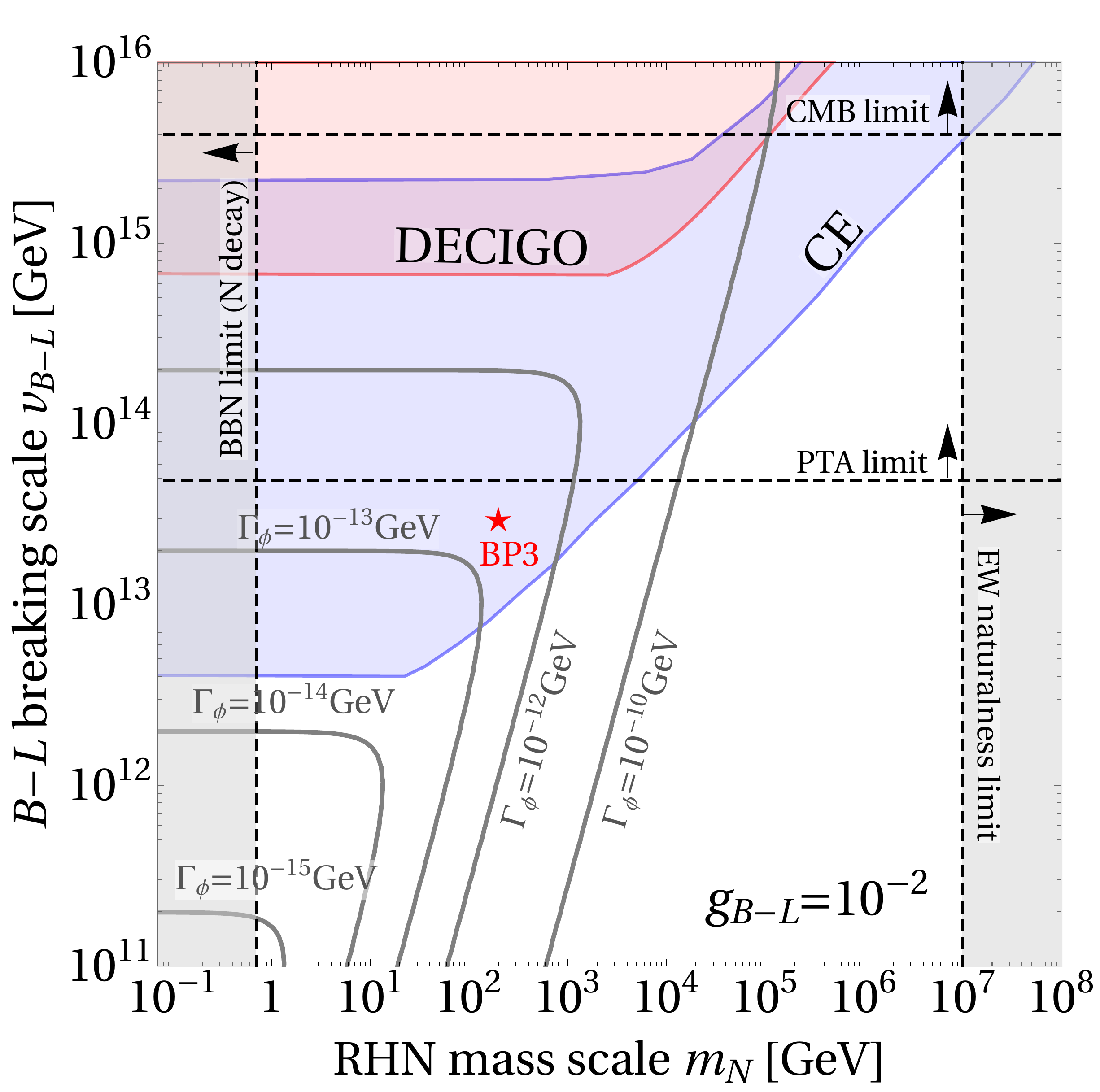}
\caption{Regions in the $m_N$--\,$v_{B-L}$ plane where the break in the GW spectrum can be observed by CE, DECIGO, and LISA; for $\lambda_\phi = 4\pi g_{B-L}^4$ and $g_{B-L} = 10^{-4}$ (\textit{left}), $10^{-3}$ (\textit{middle}), $10^{-2}$ (\textit{right}).
%
%
Whenever $\phi\rightarrow N_i N_i$ remains the dominant decay channel, arbitrary $\lambda_\phi$ values, $g_{B-L}^4/\left(16\pi^2\right)\ll \lambda_\phi \ll g_{B-L}^2$, lead to the same results as long as the horizontal axis is rescaled as $m_N \rightarrow (4\pi g_{B-L}^4/\lambda_\phi)^{1/4}\,m_N$.
This introduces an uncertainty in $m_N$ of a factor of $50$, $20$, $5$ for $g_{B-L} = 10^{-4,-3,-2}$, respectively.}
\label{fig:scan}
\end{figure*}


The Lagrangian in Eq.~\eqref{eq:L} can also be regarded as a particular, extended version of the Abelian Higgs model.
Upon spontaneous symmetry breaking, it thus results in the formation of local cosmic strings with tension~\cite{Bogomolny:1975de,Hindmarsh:1994re}
\begin{equation}
G\mu = \frac{\pi v_{B-L}^2}{8\pi M_{\rm Pl}^2}\,B\left(\beta\right) \,, \quad \beta = \frac{\lambda_\phi}{2\,g_{B-L}^2} \,,
\end{equation}
where $M_{\rm Pl} = (8\pi\,G)^{-1/2} \simeq 2.44 \times 10^{18}\,\textrm{GeV}$ denotes the reduced Planck mass, and where $B \sim 0.1$~\cite{Hill:1987qx}.
Our analysis is based on the assumption that the dynamics of cosmic strings is well described by the Nambu--Goto action, which increases the strength of the predicted GW signal (see the discussion in \cite{Auclair:2019wcv} and references therein).


In order to study the $B\!-\!L$ phase transition, we now compute the effective potential $V_{\rm eff}\left(\phi,T\right)$ (see \cite{Laine:2016hma} for a review), including the Coleman-Weinberg potential~\cite{Coleman:1973jx}, finite-temperature corrections at one-loop order~\cite{Dolan:1973qd}, and higher-order terms via a resummation of ring terms~\cite{Arnold:1992rz}.
We assume that the Universe is reheated to a large temperature $T_{\rm rh}$ after inflation, such that $U(1)_{B-L}$ is initially unbroken and the expansion driven by radiation.
At early times, the scalar field is hence stabilized at the origin by its thermal mass, $m_{\phi,\rm eff}^2\left(T\right) = \partial_\phi^2\,V_{\rm eff}\left(\phi,T\right) > 0$.
In the scenario that we are interested in, the $B\!-\!L$ phase transition is of second order.
The field $\phi$ therefore begins to evolve when its thermal mass flips sign, $m_{\phi,\rm eff}^2\left(T_{\rm c}\right) = 0$, at some critical temperature $T_{\rm c}$.
At this moment, it begins to roll down the effective potential and oscillate around the true vacuum.
After a few oscillations, the scalar-field energy density $\rho_\phi$ behaves like the energy density of ordinary matter, which allows us to model the subsequent evolution by the following set of Boltzmann equations,
\begin{align}
\label{eq:BEphi}
\dot{\rho}_\phi\left(t\right) + 3\,H\left(t\right)\rho_\phi\left(t\right) = & - \Gamma_\phi\,\rho_\phi\left(t\right) \,, \\
\label{eq:BErad}
\dot{\rho}_{\rm rad}\left(t\right) + 4\,H\left(t\right)\rho_{\rm rad}\left(t\right) = & + \Gamma_\phi\,\rho_{\rm rad}\left(t\right) \,,
\end{align}
where the Hubble rate $H$ satisfies the Friedmann equation, $3M_{\rm Pl}^2\,H^2 = \rho_\phi + \rho_{\rm rad}$.
The initial energy densities at $T_{\rm c}$ are $\rho_\phi\left(T_{\rm c}\right) = V_{\rm eff}\left(0,T_{\rm c}\right)$ and $\rho_{\rm rad}\left(T_{\rm c}\right) = \pi^2/30\,g_*T_{\rm c}^4$, where $g_* = 116$ in the minimal gauged $B\!-\!L$ model.


The solution of Eqs.~\eqref{eq:BEphi} and \eqref{eq:BErad} for BP1 is shown in Fig.~\ref{fig:BPs}, which illustrates several features of our scenario:
(i) In the parameter region of interest, $V_{\rm eff}\left(0,T\right)$ is always subdominant compared to the radiation energy density at $T \geq T_{\rm c}$.
We therefore never encounter a second period of inflation.
(ii) The onset of the scalar era always only occurs after a large number of oscillations.
There is hence no need to time-resolve the dynamics of the phase transition in more detail.
(iii) The scalar field always safely decays before big-bang nucleosynthesis (BBN)~\cite{Kawasaki:2017bqm}.


Our scenario builds upon the assumption that the scalar decay rate is much smaller than the scalar mass, $\Gamma_\phi \ll m_\phi$.
To ensure that this condition is fulfilled, we perform a careful study of all possible $\phi$ decay channels, including higher-order radiative corrections.
We are thus able to identify the following viable parameter space:
(i)~In order to suppress $\phi\rightarrow Z'Z'$ decays (on or off shell), we require $\lambda_\phi \ll g_{B-L}^2$, which makes $\phi$ parametrically lighter than the $Z'$ boson~\cite{Djouadi:2005gi}.
For definiteness, we set $\lambda_\phi = 4\pi g_{B-L}^4$, which ensures that $\lambda_\phi$ is stable against radiative corrections, which are of order $g_{B-L}^4/(16\pi^2)$.
(ii)~In order to suppress $\phi\rightarrow N_iN_i$ decays, we require small RHN Yukawa couplings, $y_N \lesssim 10^{-7}$.
In fact, this is also necessary to minimize the radiative corrections to the portal coupling $\lambda_{\phi h}$, which would otherwise induce fast $\phi\rightarrow hh$ decays.
(iii)~For the same reason, we need to choose a small portal coupling at tree level.
This choice, however, is well motivated by EW naturalness.
For simplicity, we therefore assume the $\Phi$ and $H$ sectors to be sequestered at tree level, such that the portal coupling between them is only generated at one loop, $\lambda_{\phi h} \sim (y_i^{\rm M})^2(y_{i\alpha}^{\rm D})^2/(16\pi^2)$, which is small enough to keep the radiative corrections to the Higgs mass under control.


The parameter space defined by these three conditions is radiatively stable and technically natural.
The remaining free parameters are $g_{B-L}$, $v_{B-L}$, and $m_N$. 
In Fig.~\ref{fig:scan}, we present a scan over these parameters indicating the regions for which we expect the break in the GW spectrum at frequency $f_{\rm brk}$ and amplitude $\Omega_{\rm gw}\left(f_{\rm brk}\right)$ to be within the sensitivity reach of CE, DECIGO, and LISA, respectively. 
Here, we go beyond Eq.~\eqref{eq:fbrk} and compute $f_{\rm brk}$ based on the numerical solutions of Eqs.~\eqref{eq:BEphi} and \eqref{eq:BErad} as the frequency that satisfies $t_1\left(f\right) = t_{\rm end}$~\cite{Cui:2018rwi}.
We also check that the phase transition is always of second order.
In particular, we confirm that any thermal barrier in the scalar potential that could in principle lead to a first-order phase transition is always very short-lived.


An important result of our parameter scan is that $\Gamma_\phi$ is mostly dominated by the $\phi\rightarrow N_iN_i$ partial width,
\begin{equation}
\Gamma\left(\phi\rightarrow N_i N_i\right) \simeq \frac{3\,y_N^2}{32\pi}\,m_\phi \,.
\end{equation}
In Fig.~\ref{fig:scan}, this is everywhere the case where the $\Gamma_\phi$ contours vary with the RHN mass scale $m_N$.
Only for large gauge coupling and small RHN masses, we find a different dominant decay channel\,---\,the radiative three-body decay into a SM fermion pair $f\bar{f}$ and a SM gauge boson $V = \left\{\gamma,Z,W,g\right\}$ via a $Z'Z'f$ one-loop triangle diagram,
\begin{equation}
\Gamma\left(\phi \rightarrow f\bar{f}\,V\right) \sim 10^{-8}\,\lambda_\phi\,g_{B-L}^4\,m_\phi \,.
\end{equation}
We use the software tool \textsc{Package-X}~\cite{Patel:2015tea} to confirm that this channel dominates over the corresponding $\phi\rightarrow f\bar{f}$ two-body decay, which is suppressed by a factor $m_f^2/m_{Z'}^2$ due to a chirality flip (see also~\cite{Han:2017yhy} for a similar effect).


\medskip\noindent\textbf{Conclusions.}
Leptogenesis at low and intermediate energy scales provides an attractive baryogenesis scenario that relates the origin of the matter--antimatter asymmetry to new physics in the neutrino sector and that is at the same time in accord with the concept of EW naturalness.
In this paper, we presented a unique possibility to test this scenario via observations of the stochastic GW background that originates from a network of cosmic strings after the cosmological breaking of a $U(1)_{B-L}$ gauge symmetry. 
We argued that detecting a characteristically shaped break in the GW spectrum would represent a smoking gun for a scalar era after the $B\!-\!L$ phase transition driven by a weakly coupled symmetry-breaking scalar field.
Such a detection would provide a handle on \textit{two} fundamental energy scales: the energy scale of spontaneous $B\!-\!L$ breaking and the mass scale of the RHNs that become massive during the $B\!-\!L$ phase transition.


The physical picture sketched in this paper is straightforwardly realized in the minimal gauged $B\!-\!L$ model, for which we found that a large fraction of the viable parameter space will be probed in future GW experiments (see Fig.~\ref{fig:scan}).
Our general guiding principles, however\,---\,high-scale $B\!-\!L$ breaking as motivated by grand unification and low-scale leptogenesis as motivated by EW naturalness\,---\,extend beyond this concrete model and call for further investigations of the expected GW spectrum.


\medskip\noindent\textit{Note added.}
After the completion of this work, the NANOGrav pulsar timing experiment reported on strong evidence for a new stochastic common-spectrum process in their 12.5-year data set~\cite{Arzoumanian:2020vkk}.
This signal is consistent with an interpretation in terms of nanohertz gravitational waves from a network of cosmic $B\!-\!L$ strings with a tension of around $G\mu \sim 10^{-10}$, if $\alpha = 0.1$~\cite{Ellis:2020ena,Blasi:2020mfx}.
For smaller values of $\alpha$, or assuming metastable cosmic strings, even larger values of $G\mu$ are possible~\cite{Blasi:2020mfx,Buchmuller:2020lbh}. 
If confirmed in the future, the NANOGrav signal indicates that LISA, DECIGO, and CE will have excellent chances to chart the cosmic-string-induced GW spectrum, including the characteristic features discussed in this paper.


\medskip\noindent\textit{Acknowledgments.} We thank Simon J.\ D.\ King for fruitful discussions and the organizers of the BLV2019 workshop in Madrid, where this project was initiated.
This project has received funding from the European Union's Horizon 2020 Research and Innovation Programme under grant agreement number 796961, ``AxiBAU'' (K.\,S.).


\bibliographystyle{JHEP}
\bibliography{arxiv_2}


\end{document}